\documentclass[AMA,STIX1COL]{WileyNJD-v2}
\usepackage{moreverb}

\newcommand\BibTeX{{\rmfamily B\kern-.05em \textsc{i\kern-.025em b}\kern-.08em
T\kern-.1667em\lower.7ex\hbox{E}\kern-.125emX}}

\articletype{Research Article}%

\received{31 August, 2021}
\revised{10 December, 2021}
\accepted{21 March, 2022}

\usepackage{setspace}
\usepackage{color}
\usepackage{amsmath}
\usepackage{amsfonts}
\usepackage{verbatim}
\usepackage{amsmath}
\usepackage{graphicx}
\usepackage[caption=false]{subfig}%
\providecommand{\U}[1]{\protect\rule{.1in}{.1in}}
\newcommand{\be}{\begin{equation}}
\newcommand{\ee}{\end{equation}}

\newcommand{\mincir}{\raise
-3.truept\hbox{\rlap{\hbox{$\sim$}}\raise4.truept\hbox{$<$}\ }}
\newcommand{\magcir}{\raise
-3.truept\hbox{\rlap{\hbox{$\sim$}}\raise4.truept\hbox{$>$}\ }}

\RequirePackage{latexsym}
\usepackage{hyperref}
\hypersetup{
   colorlinks=true,
    linkcolor=red,
   filecolor=magenta,      
    urlcolor=cyan,
    pdfpagemode=FullScreen}

\usepackage{tikz,xcolor,hyperref}

\definecolor{lime}{HTML}{A6CE39}
\DeclareRobustCommand{\orcidicon}{%
	\begin{tikzpicture}
	\draw[lime, fill=lime] (0,0) 
	circle [radius=0.16] 
	node[white] {{\fontfamily{qag}\selectfont \tiny ID}};
	\draw[white, fill=white] (-0.0625,0.095) 
	circle [radius=0.007];
	\end{tikzpicture}
	\hspace{-2mm}
}

\foreach \x in {A, ..., Z}{%
	\expandafter\xdef\csname orcid\x\endcsname{\noexpand\href{https://orcid.org/\csname orcidauthor\x\endcsname}{\noexpand\orcidicon}}
}

\begin{document}

\title{Lie Symmetries, Painlev\'{e} analysis and global dynamics for the temporal
equation of radiating stars}

\author[1,2]{Genly Leon*}

\author[3]{Megandhren Govender}

\author[2,4]{Andronikos Paliathanasis}

\authormark{GENLY LEON \textsc{et al}}

\address[1]{Departamento de Matem\'aticas, Universidad Cat\'olica del Norte, Avda. Angamos 0610, Casilla 1280 Antofagasta, Chile}

\address[2]{Institute of Systems Science, Durban University of Technology, Durban 4000,
South Africa}

\address[3]{Department of Mathematics, Faculty of Applied Sciences, Durban University of
Technology, Durban 4000, South Africa}

\address[4]{Instituto de Ciencias F\'{\i}sicas y Matem\'{a}ticas, Universidad Austral de
Chile, Valdivia 5090000, Chile}

\corres{*Genly Leon, Departamento de Matem\'aticas, Universidad Cat\'olica del Norte, Avda. Angamos 0610, Casilla 1280 Antofagasta, Chile. \email{genly.leon@ucn.cl}}

%\presentaddress{Present address}

\abstract[Abstract]{We study the temporal equation of radiating stars by using three powerful methods for the analysis of nonlinear differential equations. Specifically, we investigate the global dynamics for the given master ordinary differential equation to understand the evolution of solutions for various initial conditions as also to investigate the existence of asymptotic solutions. Moreover, with the application of Lie's theory, we can reduce the order of the master differential equation, while an exact similarity solution is determined. Finally, the master equation possesses the Painlev\'{e} property, which means that the analytic solution can be expressed in terms of a Laurent expansion.}

\doi{10.1002/mma.8274}


\jnlcitation{Leon G, Govender M, Paliathanasis A. Lie symmetries, Painlevé analysis, and global
dynamics for the temporal equation of radiating stars. {\it{Math Meth Appl Sci.}} 2022;1-16. doi:10.1002/mma.8274}

\maketitle

\section{Introduction}

Ordinary differential equations play an important role in the study of physical systems. There are various approaches for the study of the properties of physical systems described by differential equations as well as to construct and determine exact and analytic solutions. Relativistic astrophysics is an active area employing various solution-generating techniques to solve highly nonlinear systems of differential equations. From finding exact solutions of the Einstein field equations or their modifications (Einstein-Gauss Bonnet gravity, $f(R)$ gravity or Brans-Dicke theory, etc.) describing stellar objects through to the study of causal thermodynamics, where solutions of governing differential equations have played a key role. 

The end-states of gravitational collapse of bounded configurations have held the attention of astrophysicists since the pioneering work of Oppenheimer and Snyder\cite{oppen} in which they studied idealised collapse of a dust sphere. The Weak Cosmic Censorship Conjecture, first articulated by Penrose, forbids the existence of naked singularities arising from continued gravitational collapse. However, there have been many counter-examples put forward within the framework of Einstein's classical general relativity \cite{j1,ong,j2,wagh}. The confirmation of classical general relativity as a cornerstone of gravitational theory was borne out in 2019 when the photograph of the shadow of a black hole \cite{shadow} was obtained, heralding a new frontier of theoretical predictions and observations. The discovery of the Vaidya solution \cite{vaidya} paved the way for researchers to study dissipative collapse of stars.  The boundary of the collapsing object divides spacetime into two distinct regions, ${\cal M^-}$, the interior region and ${\cal M^+}$, the exterior region described by the Vaidya solution.  Early work on dissipative gravitational collapse can be attributed to Herrera and co-workers (see  \cite{h1,h2} and references therein). In their investigations they studied spherically symmetric, shear-free stellar objects undergoing dissipative gravitational collapse in the form of a radial heat flux. The junction conditions for the smooth matching of the interior spacetime to the exterior Vaidya solution were derived by Santos  \cite{santos}. The Santos junction conditions demonstrated that the pressure at the boundary of a collapsing, radiating star is nonzero and is proportional to the magnitude of the outgoing radial heat flux. This junction condition represents the conservation of momentum across the boundary of the collapsing star. Recently, the Santos junction conditions have been extended to include a dynamically unstable core with a general energy-momentum tensor describing an imperfect fluid with heat flux and null radiation with the exterior being described by the generalised Vaidya solution \cite{sunilbrassels}. Over the next two decades, the study of radiating stars has provided us with a rich insight into the end-states of continued gravitational collapse, particularly with regards to time of formation of the horizon\cite{nolene}, temperature profiles and relaxational effects related to causal heat flux\cite{Tron}. The shear-free models were subsequently extended to include the effects of shear viscosity. It has been demonstrated that the Chandrasekhar stability criterion for isotropic fluid spheres is modified in the presence of shear viscosity. Furthermore, in both the Newtonian and relativistic regimes, the shear viscosity decreases the instability of the stellar fluid\cite{chano}.

The inclusion of shear, anisotropy, electromagnetic field, and rotation in the slow approximation have been fruitful in studying the thermodynamics of such systems. The impact of shear on the kinematics and dynamics of the collapse process has been addressed by several authors. The instability of the shear-free condition has been demonstrated by Herrera et al. \cite{instability1} in which they showed that the shear-free condition may hold for a limited epoch of the collapse process. The presence of pressure anisotropies, density inhomogeneities, and dissipative fluxes can mimic shear-like effects \cite{mimic}. The inclusion of shear during dissipative collapse has led to interesting results when contrasted to the shear-free case. Shearing effects lead to higher core temperatures it has been shown that horizon formation is delayed when shear is present. In an attempt to model shearing, radiating stars, Ivanov introduced the so-called {\it horizon function} which simplifies the boundary condition representing the temporal behavior of the model\cite{ivanova,ivanovb}. The horizon function has a physical attribute in the sense that it is directly related to the surface redshift of the collapsing sphere. Once this function is determined from the boundary condition, the end-state or possible outcome of continued gravitational collapse can be studied. The avoidance of the singularity was demonstrated by using a simple model of shear-free collapse in which the rate of collapse is balanced by the rate of energy emission to the exterior spacetime \cite{bcd}. This so-called horizon-free model, or Banerjee, Chatterjee \& Dadhich (BCD) \cite{bcd} model, forms the basis of the work contained in this paper. The horizon-free collapse in the presence of shear was studied in various models in which the gravitational potentials were highly simplified. The so-called Euclidean stars, in which the areal radius is equal to the proper radius were shown to undergo horizon-free collapse \cite{euclid}. The BCD model has two inherent assumptions: the interior spacetime is shear-free and the gravitational potentials are separable in space and time. Works by Chan and co-workers attempted to generalise the shear-free model to include shear by demanding that the metric functions be separable. These models have several drawbacks such as the proper radius being independent of time or the resulting boundary condition rendered solvable via numerical methods\cite{chanr1,chanr2}.    

In this piece of work, we investigate the dynamics for the master equation of
the temporal equation of radiating stars. The equation that we are interested
describes the Einstein field equations for a spherically symmetric line
element%
\begin{equation}
ds^{2}=-A^{2}(r,t)dt^{2}+B^{2}(r,t)\left(  dr^{2}+r^{2}d\theta^{2}+r^{2}%
\sin^{2}\theta d\phi^{2}\right),
\end{equation}
for a pressure isotropy fluid with an exterior solution the Vaidya metric
 \cite{vaidya} such that to describe a radiative solution. Banerjee et al.
 \cite{bhui} suggested the metric ansatz $A\left(  t,r\right)  =1+\zeta
_{0}r^{2},~B\left(  t,r\right)  =R(t)~$where $\zeta_{0}$ is a positive
constant. In this case, the field equations reduce to the second-order
differential equation
\begin{equation}
2R(t){\ddot{R}}(t)+{\dot{R}(t)}^{2}+\alpha{\dot{R}(t)}=\beta\label{bc},%
\end{equation}
where $\alpha$ and $\beta$ are constants and the dot means derivative with respect to the time $t$. A special solution of the latter equation is the solution $R=-Ct$ where $C>0$ is a constant, which describes a
collapsing star  \cite{bhui}. Recently, in  \cite{andromegan} new families of
exact solutions for the master equation (\ref{bc}) were found using Lie symmetries. This body of work forms part of several investigations about dissipative collapse via Lie symmetries. Radiating stars with shear, in which the particle trajectories within the stellar fluid were geodesics, were studied using Lie symmetries\cite{liea}. Several new solutions were obtained while other solutions were reduced to well-known cases studied earlier in the literature. The geodesic case with shear was extended to the most general shearing matter distribution radiating energy to the exterior spacetime. The method of Lie symmetries proved useful in obtaining four new classes of solutions \cite{lieb}, two of which included the horizon function and the Euclidean condition. 

In the
following, we study the dynamics and the stability properties for the master
equation (\ref{bc}) and also, we show in detail how the method of Lie point
symmetries and the singularity analysis can be used for the derivation of new
analytic and exact solution for equation (\ref{bc}). 

The theory of Lie symmetries  \cite{kumei} is a powerful approach for the
investigation of invariant functions for differential equations.\ The novelty
of Lie theory is that the infinitesimal representations of the finite
transformations of continuous groups are considered, by moving from the group
to a local algebraic representation, and to studying the invariance properties
under them. The resulting invariant functions can be used for the reduction of
order for a given ordinary differential equation and consequently for the
derivation of solutions. These exact solutions which follow from the
application of Lie symmetries are known as similarity solutions. The theory of
Lie symmetries cover a range of applications in physical science and 
gravitational theory, see for instance  \cite{sm1,sm2,sm3,sm4,sm5,sm6} and
references therein.

The singularity analysis, which is mainly associated
with the school of Painlev\'{e}, that is  why also it is called Painlev\'{e}
analysis  \cite{pain1}, is another powerful mathematical approach for the derivation of solutions of
differential equations. When a differential equation passes the singularity
the analysis we say that the differential equation possesses the Painlev\'{e}
property. Nowadays the application of the singularity analysis is summarised
in ARS algorithm, from the initials of Ablowitz, Ramani, and Segur
 \cite{ars1,ars2,ars3} who established a systematic method for investigation of
analytic solutions, inspired by the approach applied by Kowalevskaya
 \cite{Kowalevski88} for the determination of the third integrable case of
Euler's equations for a spinning top. The basic characteristic for 
a differential equation to possesses the Painlev\'{e} property is the
existence of at least a movable singularity. In the singularity analysis the
analytic solution for a given differential equation is expressed by
Painlev\'{e} Series, and specifically in our consideration we shall write the analytic solution in the Puiseux series.

On the other hand, stability analysis provides us with an important tool to investigate 
the evolution of the given dynamical system. Indeed, from the stability
analysis, we can investigate if a given solution is stable or not, while we can
determine families of initial conditions such that specific behavior for the
dynamical system to be stable. In addition, we can define constraints for the
free parameters of the differential equations according to the stability of
exact solutions. Hence, we can extract important information about the nature
of the free parameters. In our consideration, for equation (\ref{bc}) \ we can
understand how the free parameters $\alpha$ and $\beta$ affects the dynamics.
The plan of the paper is as follows.

In Section \ref{sec2} we investigate the stability properties for the master
equation (\ref{bc}) for the exact solution found in \cite{bhui}, while we perform a
detailed study of the global dynamics for the master equation. From the
dynamical analysis, a new family of exact solutions is constructed. In Section
\ref{sec3}, we investigate equation (\ref{bc}) by applying Lie's theory, as we
study if equation (\ref{bc}) possesses the Painlev\'{e} property. We find that
the differential equation admits two Lie point symmetries which form the
$A_{2,2}$ Lie algebra. Consequently, the application of the Lie symmetries
provides two similarity transformations. Moreover, the application of the ARS
algorithm indicates that the master equation (\ref{bc}) satisfies the
Painlev\'{e} test, and the analytic solution is expressed by a right
Painlev\'{e} series. However, for a specific relation between the two free
variables $\alpha$, $\beta$; a new exact solution is determined. Finally, in
Section \ref{sec4} we discuss our results and draw our conclusions.

\bigskip

\section{Stability analysis and global dynamics}

\label{sec2}

For the master equation (\ref{bc}), \ by convenience we assume $\alpha>0$. A
special and simple solution to this equation is $R=-Ct$ defined for
$-\infty<t\leq0$, where $C>0$ is a constant given by the positive roots of
\begin{equation}
    -\beta+C^{2}-\alpha C=0.
\end{equation} That is, for $\beta>0$, we have only one positive root $C=\frac{1}{2}\left(
\sqrt{\alpha^{2}+4\beta}+\alpha\right)$. For $\beta<0$ and $\alpha^{2}+4\beta>0$, we have two different positive
roots:  $C_{-}=\frac{1}{2}\left(  \alpha-\sqrt{\alpha^{2}+4\beta
}\right)$ and $\quad C_{+}=\frac{1}{2}\left(  \sqrt{\alpha^{2}+4\beta}
+\alpha\right)  $.

To avoid ambiguities we prefer to use the relation $\beta= C^{2} - C \alpha$, and in the analysis separate the cases $\beta<0$ and $\beta\geq0$.

For the analysis of stability of the scaling solution $R(t)=-C t$, with
$C=\frac{1}{2} \left(  \sqrt{\alpha^{2}+4 \beta}+\alpha\right)  >0$ in the
interval $-\infty< t <0$ we use similar methods as in Liddle \& Scherrer
 \cite{Liddle:1998xm} and Uzan  \cite{Uzan:1999ch}. For this purpose, the new logarithmic time variable $\tau$ related to $t$ through
\begin{equation}
\label{t-tau-transform}
t= -e^{-\tau}, \quad  -\infty <\tau <\infty,
\end{equation}
such that $t\rightarrow -\infty$ as $\tau \rightarrow -\infty$ and  $t\rightarrow 0$ as $\tau \rightarrow +\infty$ is defined. 

Then, replacing the relation between $t$ and $\tau$ given by \eqref{t-tau-transform} in the expression $R(t)$, we obtain a function of $\tau$ from $R(t)$ denoted and defined by 
\begin{equation}
    \bar{R}(\tau)= R(-e^{-\tau}). 
\end{equation}
Using again the transform \eqref{t-tau-transform}, the scaling solution $R (t)=-C t$ can be re-expressed as a function of $\tau$ as $R_s(\tau)= C e^{-\tau}$.

To compare a generic solution  $\bar{R}(\tau)$ with the scaling solution $R_s(\tau)= C e^{-\tau}$, we define a dimensionless function of $\tau$ as the 
ratio 
\begin{equation}
\label{def_u}
u(\tau)= \frac{\bar{R}(\tau)}{R_s(\tau)}.
\end{equation}
Thus, the solution $R_s(\tau)$ will corresponds to the equilibrium point $u=1$ in the new formulation of the  master equation (\ref{bc}). 
The physical region corresponds to $u\geq0$. Recall, the physical solution $R = -Ct$ is defined
for $-\infty< t \leq0$, where $C>0$ is a constant.

To reformulate the master equation (\ref{bc}) in terms of $u(\tau)$, and its derivatives with respect to $\tau$,  we use the chain rule to obtain several differentiation rules. 
Let be denoted the time derivative with respect to $\tau$ be denoted by a prime and the time derivative with respect to $t$ by a dot. That is, for $f(\tau)$, let 
\begin{equation}
f'(\tau) \equiv \frac{d f(\tau)}{d \tau},
\end{equation}
and for $g(t)$, let 
\begin{equation}
\dot{g}(t) \equiv \frac{d g(t)}{d t}.
\end{equation}
Then, 
\begin{equation}\label{rules}
\dot R (t) = \frac{d \tau (t)}{d t} \bar{R}^{\prime}(\tau) = e^{\tau} \bar{R}'(\tau), \quad 
\ddot R(t) = e^{2 \tau} \left(\bar{R}^{\dprime}(\tau) + \bar{R}^{\prime}(\tau) \right)
\end{equation}
and 
\begin{equation}
{R_s^{\prime}(\tau)}/{R_s(\tau)}= -1, \quad R_s(\tau)= C e^{-\tau}.
\end{equation}
Therefore, by using the rules \eqref{rules}, and substituting expression \eqref{t-tau-transform}, the equation (\ref{bc}) becomes 
\begin{equation}
-\beta +\alpha  e^{\tau } \bar{R}^{\prime}(\tau)+e^{2 \tau } \left({\bar{R}^{\prime}(\tau)}^2+2 \bar{R}(\tau) \left(\bar{R}^{\dprime}(\tau)+\bar{R}^{\prime}(\tau)\right)\right)=0. \label{new1}   
\end{equation}
The next step is to rewrite equation \eqref{new1} in terms of $u(\tau)$ and its derivatives. 

Solving equation \eqref{def_u} for $\bar{R}$, and then using the derivatives rules for products of functions and the chain rule, we obtain, subsequently, 
\begin{eqnarray}
&& \bar{R}(\tau)= C e^{-\tau } u(\tau), \label{R}\\
&& \bar{R}^{\prime}(\tau)=  C e^{-\tau }
\left(u^{\prime}(\tau)-u(\tau)\right), \label{dR}\\
&& \bar{R}^{\dprime}(\tau)= C e^{-\tau } \left(u^{\dprime}(\tau)-2 u^{\prime}(\tau)+u(\tau)\right). \label{ddR}
\end{eqnarray}
Substituting \eqref{R}, \eqref{dR} and \eqref{ddR} in equation (\ref{new1}), and dividing by the overall  factor $e^{-\tau }$, we obtain 
\begin{eqnarray}
&& 2 C^2 u(\tau) u^{\dprime}(\tau)+C^2 {u^{\prime}(\tau)}^2    +C u^{\prime}(\tau) (\alpha -4 C u(\tau))  +C (u(\tau)-1) (-\alpha +C u(\tau)+C)=0. 
\end{eqnarray}
where we have used the relation 
$\beta= C^2 - C \alpha$.

Defining the new function
\begin{equation}
v(\tau)= u^{\prime}(\tau),
\end{equation}
we obtain the dynamical system
\begin{align}
&  u^{\prime}(\tau)=v(\tau),\label{equ}\\
&  v^{\prime}(\tau)=-\frac{(u(\tau)-1) (-\alpha+C u(\tau)+C)}{2 C u(\tau)} +v(\tau) \left(  2-\frac{\alpha
}{2 C u(\tau)}\right)  -\frac{v(\tau)^{2}}{2 u(\tau)}. \label{eqv}%
\end{align}
The scaling solution $R_s(\tau)= C e^{-\tau}$ corresponds to the equilibrium point $u=1, v=0$.

Defining $\varepsilon$ through $u=1+\varepsilon$, we obtain the final
dynamical system 
\begin{align}
&  \varepsilon^{\prime}(\tau)= v(\tau),\\
&  v^{\prime}(\tau)=\frac{\varepsilon(\tau)(\alpha-C \varepsilon(\tau)-2 C)}{2 C (\varepsilon(\tau)+1)}+v(\tau)
\left(  2-\frac{\alpha}{2 C \varepsilon(\tau)+2 C}\right)  -\frac{v(\tau)^{2}}{2
(\varepsilon(\tau)+1)},
\end{align}
where the equilibrium point is translated to the origin.
\begin{figure}[ptb]
\centering
\includegraphics[width=0.8\textwidth]{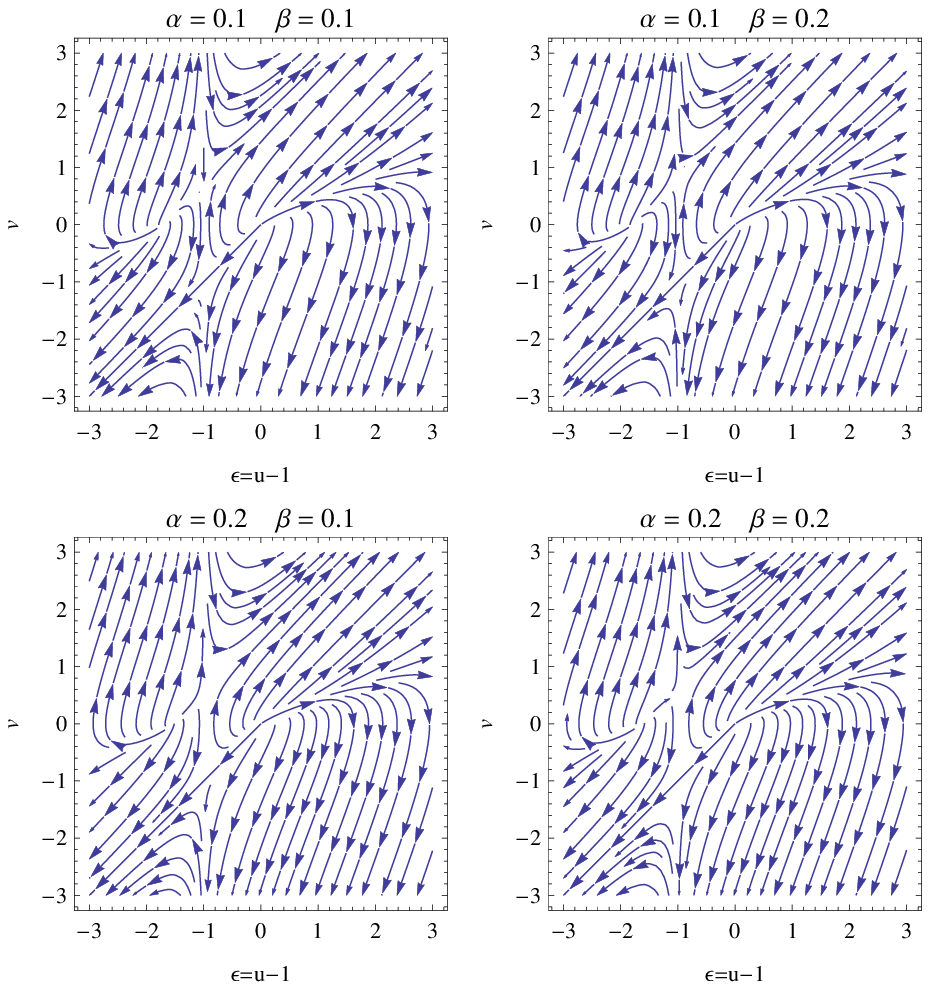} \caption{Phase plot of
(\ref{equ}), (\ref{eqv}) for some $\alpha>0,\beta>0$ and the choice
$2C:=\alpha+\sqrt{\alpha^{2}+4\beta}$. The origin represents the solution
  $R_{s}(\tau)=Ce^{-\tau}$, which is unstable (node). The
physical region corresponds to $u\geq0$.}%
\label{fig:my_label1}%
\end{figure}

\begin{figure}[ptb]
\centering
\includegraphics[width=0.8\textwidth]{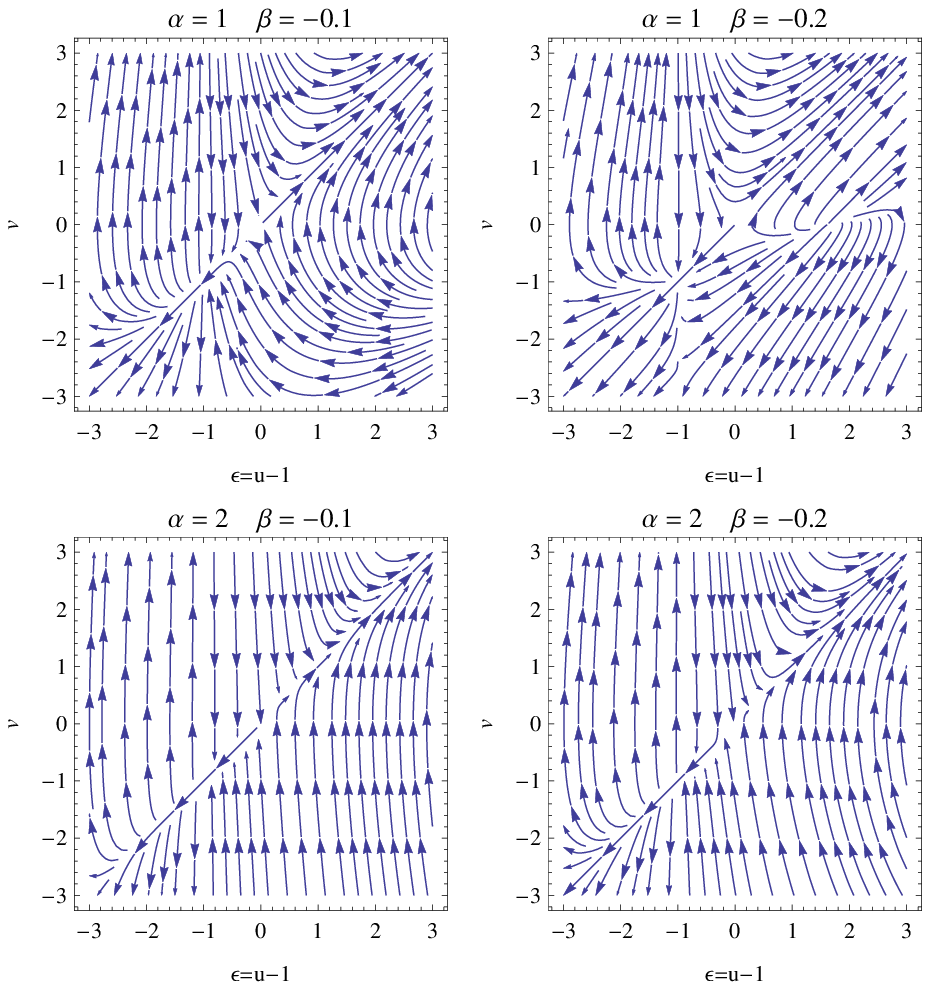} \caption{Phase plot of
(\ref{equ}), (\ref{eqv}) for some $\alpha>0,\beta<0$ and the choice
$2C_{-}:=\alpha-\sqrt{\alpha^{2}-4|\beta|}$. The origin represents the
solution  $R_{s-}(\tau)=C_{-}e^{-\tau}$, which is
unstable (saddle). The physical region corresponds to $u\geq0$.}%
\label{fig:my_label2}%
\end{figure}

\begin{figure}[ptb]
\centering
\includegraphics[width=0.8\textwidth]{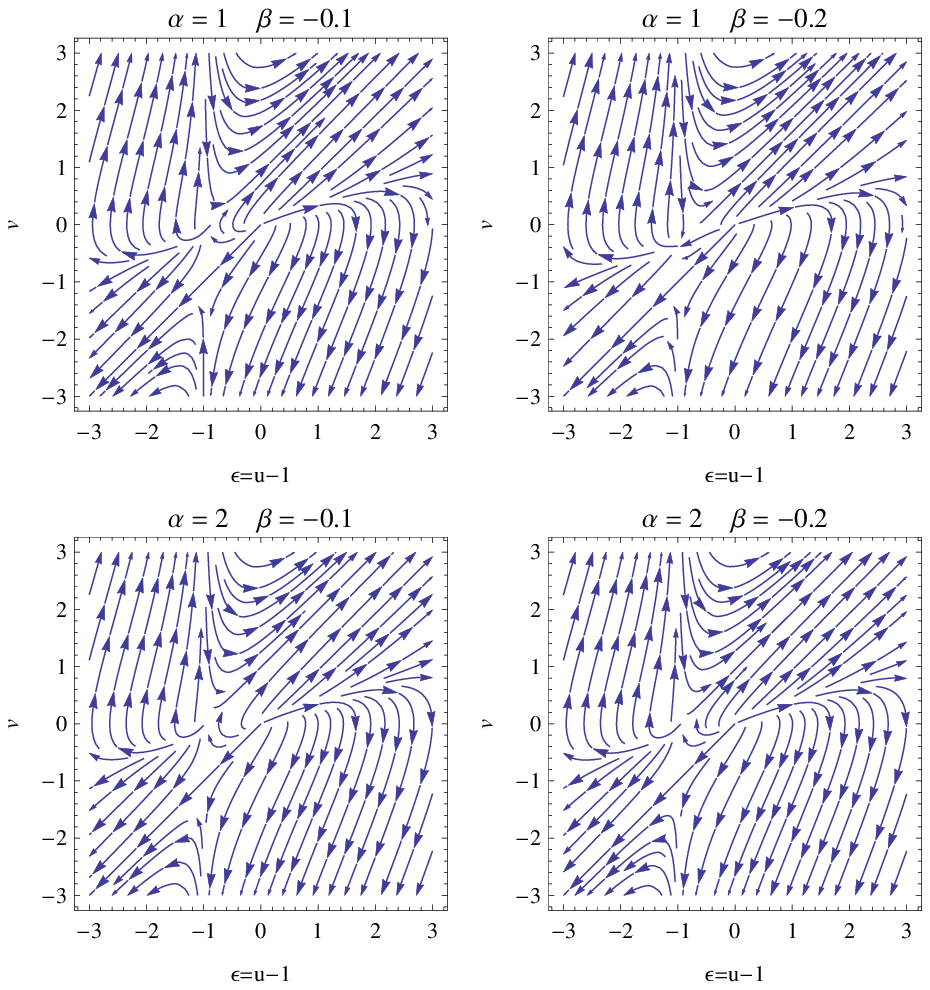} \caption{Phase plot of
(\ref{equ}), (\ref{eqv}) for some $\alpha>0, \beta<0$ and the choice $2
C_{+}:=\alpha+ \sqrt{\alpha^{2} - 4 |\beta|}$. The origin represents the
solution $R_{s+}(\tau)= C_{+} e^{-\tau}$, which is
unstable (node). The physical region corresponds to $u\geq0$.}%
\label{fig:my_label3}%
\end{figure}

The linearisation matrix is defined by 
\begin{equation}
J(\varepsilon, v)= \left(
\begin{array}
[c]{cc}%
0 & 1\\
\frac{1}{2} \left(  \frac{(v+1) (v C-C+\alpha)}{C (\varepsilon+1)^{2}}-1\right)
& 2-\frac{\alpha+2 C v}{2 \varepsilon C+2 C}
\end{array}
\right).
\end{equation}

Then, linearising around the equilibrium point, $\varepsilon=0, v=0$, we obtain
\begin{equation}
\left(
\begin{array}
[c]{c}%
\varepsilon^{\prime}(\tau)\\
v^{\prime}(\tau)%
\end{array}
\right)  = \left(
\begin{array}
[c]{cc}%
0 & 1\\
-1+ \frac{\alpha}{2 C} & 2 -\frac{\alpha}{2 C}%
\end{array}
\right)  \left(
\begin{array}
[c]{c}%
\varepsilon(\tau)\\
v(\tau)
\end{array}
\right).
\end{equation}

The linearisation matrix
\begin{equation}
J(0,0)= \left(
\begin{array}
[c]{cc}%
0 & 1\\
\frac{\alpha}{2 C}-1 & 2-\frac{\alpha}{2 C}
\end{array}
\right)
\end{equation}
has eigenvalues $\left\{  1,1-\frac{\alpha}{2 C}\right\}  $.

Assume first $\beta\geq0$. In this case, the origin is always unstable as
$\tau\rightarrow+\infty$ due to $2 C= \alpha+\sqrt{\alpha^{2} + 4\beta}%
>\alpha$. That is, the origin is stable as $\tau\rightarrow-\infty$.

An additional equilibrium point is
\begin{equation}
\varepsilon=\frac{\alpha}{C}-2<0, \quad u=\frac{\alpha}{C} -1, \quad v=0.
\end{equation}
Evaluating the linearisation matrix $J\left(\frac{\alpha}{C}-2,0\right)$, we obtain the eigenvalues $\left\{  1,\frac{2
C-\alpha}{2 (C-\alpha)}\right\}  $. Due to $2 C-\alpha>0$ it is unstable as
$\tau\rightarrow\infty$. Indeed for $0<\frac{\alpha}{2}<C <\alpha$ it is a
saddle, whereas for $C>\alpha>0$ is an unstable node. The last conditions is
forbidden due to the physical condition $u\geq0$ evaluated at $(u,v)=\left(\frac
{\alpha}{C} -1,0\right)$ implies $\alpha\geq C$.

Now, let us study the case $\beta<0, \alpha<-2 \sqrt{-\beta}$ or $\beta<0,
\alpha>2 \sqrt{-\beta} $. Henceforth, we have two solutions
\begin{equation}
R_{s \pm}(\tau)= -C_{\pm}t(\tau)= C_{\pm} e^{-\tau},
\end{equation}
where
\begin{equation}
2 C_{\pm} = \alpha\pm\sqrt{\alpha^{2} - 4|\beta|}.
\end{equation}
Observe that $2 C_{+} > \alpha$, implies that $R_{s +}(\tau)$ is an unstable
solution (unstable node) as $\tau\rightarrow\infty$. Due to $\alpha> 2 C_{-} >
0$, $R_{s -}(\tau)$ is an unstable (saddle) solution.

Figure \ref{fig:my_label1} shows the phase plot of system (\ref{equ}), (\ref{eqv})
for some $\alpha>0, \beta>0$ and the choice $2 C:=\alpha+ \sqrt{\alpha^{2} + 4
\beta}$. The origin represents the solution  $R_{s}(\tau)= C e^{-\tau}$, which is unstable (node). 

Figure \ref{fig:my_label2} shows the phase plot of system (\ref{equ}), (\ref{eqv})
for some $\alpha>0, \beta<0$ and the choice $2 C_{-}:=\alpha- \sqrt{\alpha^{2}
- 4 |\beta|}$. The origin represents the solution 
$R_{s-}(\tau)= C_{-} e^{-\tau}$, which is unstable (saddle). 

Figure \ref{fig:my_label3} shows the phase plot of system (\ref{equ}), (\ref{eqv})
for some $\alpha>0, \beta<0, \alpha^{2}+4 \beta\geq0$ and the choice $2
C_{+}:=\alpha+ \sqrt{\alpha^{2} - 4 |\beta|}$. The origin represents the
solution  $R_{s+}(\tau)= C_{+} e^{-\tau}$, which is
unstable (node).

As can be seen in figures \ref{fig:my_label1}, \ref{fig:my_label2} and
\ref{fig:my_label3}, there are non-trivial dynamics as $(u,v)$ are unbounded.

\subsection{Dynamics as $(u,v)$ are unbounded}

Assume that there are $u_{0}> 0$, and a coordinate transformation $\phi=h(u)$,
with inverse $h^{(-1)}(\phi)$, which maps the interval $[u_{0},\infty)$ onto
$(0, \delta]$, where $\delta=h(u_{0})$, satisfying $\lim
_{u\rightarrow+\infty}h(u)=0$, and has the following additional properties:

\begin{enumerate}
\item $h$ is $C^{k+1}$ and strictly decreasing,

\item
\begin{equation}
\overline{h'}(\phi)=\left\{
\begin{array}
[c]{cc}%
h^{\prime}(h^{(-1)}(\phi)), & \phi>0,\\
\lim_{\phi\rightarrow\infty} h'(\phi), & \phi=0
\end{array}
\right.  \label{eq23}%
\end{equation}
is $C^{k}$ on the closed interval $[0, \delta]$ and

\item $\frac{d \overline{h'}}{d \phi}(0)$ and higher derivatives
$\frac{d^{m}\overline{h'}}{d \phi^{m}}(0)$ satisfy
\begin{equation}
\frac{d \overline{h'}}{d \phi}(0)=\frac{d^{m}\overline{h'}}{d \phi^{m}%
}(0)=0.
\end{equation}

\end{enumerate}

It can be proved using the above conditions that
\begin{align}
&  \lim_{\phi\rightarrow0}\frac{1}{h^{(-1)}(\phi)}=0,\\
&  \lim_{\phi\rightarrow0}\frac{h'\left(  h^{(-1)}(\phi)\right)
}{\phi}=0,\\
&  \lim_{\phi\rightarrow0}\quad\frac{h^{\prime\prime}\left(  h^{(-1)}%
(\phi)\right)  }{h'\left(  h^{(-1)}(\phi)\right)  }=0.
\end{align}
In the following, we say that $g$ is well-behaved at infinity (WBI) of
exponential order $N$, if there is $N$ such that
\begin{equation}
\lim_{u\rightarrow\infty} \left(  \frac{g'(u)}{g(u)}-N\right)  =0.
\end{equation}

Let $g$ be a WBI function of exponential order $N$ then, exponential dominated
means, for all $\lambda>N$,
\begin{equation}
\lim_{ u \to\infty} \, e^{-\lambda u} g(u)=0.
\end{equation}
From
\begin{equation}
\lim_{\phi\to0} \, \frac{h^{\prime\prime}\left(  h^{(-1)}(\phi)\right)
}{h'\left(  h^{(-1)}(\phi)\right)  }=0,
\end{equation}
it follows that $g(u)=1/h'(u)$ is WBI of exponential order $0$, that
is, $\lim_{u\rightarrow\infty} \frac{g'(u)}{g(u)}-N=0$ for $N=0$, and
hence it is exponential dominated. This implies in turn that $1/h(u)$ is also
exponential dominated. The function $h(u)$ must therefore obey the following
condition: for all $k>0$,
\begin{equation}
\lim_{u \rightarrow\infty} \frac{e^{k u}}{h'\left(  u\right)  }=
\lim_{u \rightarrow\infty} \frac{e^{k u}}{h(u)}=0.
\end{equation}
In general, we can obtain functions $\phi= h(u)$ satisfying the above
conditions 1, 2, 3, and previously commented facts if we demand the existence
of $n>1$ such that the functions
\begin{equation}
\frac{1}{h^{(-1)}(\phi)}, \quad\frac{h'\left(  h^{(-1)}(\phi)\right)
}{\phi}, \quad\frac{h^{\prime\prime}\left(  h^{(-1)}(\phi)\right)  }%
{h'\left(  h^{(-1)}(\phi)\right)  },
\end{equation}
behaves as $\mathcal{O}(\phi^{n})$, and
\begin{equation}
h^{(m)}\left(  h^{(-1)}(\phi)\right)  \sim\mathcal{O}(\phi^{(m n+1)}), \quad m
\in\mathbb{N}, \quad m \geq1,
\end{equation}
as $\phi\rightarrow0$, where the superscript $(m)$ means $m$-th derivative
with respect the argument.

Let be defined
\begin{equation}
\theta=1-u+v.
\end{equation}
Then we obtain
\begin{align}
&  \phi'= \overline{h'}(\phi) \left(  h^{(-1)}(\phi)+\theta
-1\right)  ,\label{eq28}\\
&  \theta'= \frac{(2 C-\alpha) \theta}{2 C h^{(-1)}(\phi)}%
-\frac{\theta^{2}}{2 h^{(-1)}(\phi)}, \label{eq29}%
\end{align}
where $\overline{h'}(\phi)$ is defined in (\ref{eq23}) and $2 C-\alpha>0$.
The system (\ref{eq28}), (\ref{eq29}) defines a flow in the phase region
\begin{equation}
\Omega_{\delta}:=\left\{  (\phi, \theta)\in\mathbb{R}^{2}: 0<\phi<
h(\delta^{-1}), \theta\in K\right\}  ,
\end{equation}
where $K$ is a compact set, such that $\Omega_{\delta}$ is a positive
invariant set for large $\tau$.

The system (\ref{eq28}), (\ref{eq29}) admits a curve of equilibrium points $L: (\phi, \theta)=
(0, \theta^{*})$ ~ parametrised by $\theta^{*}$ that is approached as $\tau\rightarrow\infty$ (for bounded $\theta$). This curve of equilibrium points is represented by a red dashed line in figures \ref{fig:my_label4}, \ref{fig:my_label5} and \ref{fig:my_label6}. 

The linearisation matrix of system (\ref{eq28}) and (\ref{eq29}) at a generic point $(\phi, \theta)$ is
\begin{align}
 J(\phi,\theta)&   = \left(
\begin{array}
[c]{cc}%
1+\frac{\left(  \theta+h^{(-1)}(\phi)-1\right)  h^{\prime\prime}\left(
h^{(-1)}(\phi)\right)  }{h'\left(  h^{(-1)}(\phi)\right)  } &
h'\left(  h^{(-1)}(\phi)\right) \\
\frac{(\alpha+C (\theta-2)) \theta}{2 C h^{(-1)}(\phi)^{2} h'\left(
h^{(-1)}(\phi)\right)  } & -\frac{\alpha+2 C (\theta-1)}{2 C h^{(-1)}(\phi)}\\
&
\end{array}
\right) \nonumber\\
&  = \left(
\begin{array}
[c]{cc}%
1 + \frac{h^{(-1)}(\phi) h^{\prime\prime}\left(  h^{(-1)}(\phi)\right)
}{h'\left(  h^{(-1)}(\phi)\right)  }+\mathcal{O}(\phi^{n}) &
h'\left(  h^{(-1)}(\phi)\right) \\
\frac{(\alpha+C (\theta-2)) \theta}{2 C h^{(-1)}(\phi)^{2} h'\left(
h^{(-1)}(\phi)\right)  } & \mathcal{O}(\phi^{n})\\
&
\end{array}
\right) \nonumber\\
&  = \left(
\begin{array}
[c]{cc}%
1 + \frac{h^{(-1)}(\phi) h^{\prime\prime}\left(  h^{(-1)}(\phi)\right)
}{h'\left(  h^{(-1)}(\phi)\right)  }+\mathcal{O}(\phi^{n}) &
\mathcal{O}(\phi^{n+1})\\
\mathcal{O}(\phi^{n-1}) & \mathcal{O}(\phi^{n})\\
&
\end{array}
\right) \nonumber\\
&  \sim\left(
\begin{array}
[c]{cc}%
1+\frac{h^{(-1)}(\phi) h^{\prime\prime}\left(  h^{(-1)}(\phi)\right)
}{h'\left(  h^{(-1)}(\phi)\right)  } & 0\\
0 & 0\\
&
\end{array}
\right)  , \quad\text{as}\; \phi\rightarrow0,
\end{align}
We have,
\begin{align}
&  J(\phi,\theta)\nonumber\\
&  = \left(
\begin{array}
[c]{cc}%
1 + \frac{h^{(-1)}(\phi) h^{\prime\prime}\left(  h^{(-1)}(\phi)\right)
}{h'\left(  h^{(-1)}(\phi)\right)  }+\mathcal{O}(\phi^{n}) &
h'\left(  h^{(-1)}(\phi)\right) \\
\frac{(\alpha+C (\theta-2)) \theta}{2 C h^{(-1)}(\phi)^{2} h'\left(
h^{(-1)}(\phi)\right)  } & \mathcal{O}(\phi^{n})\\
&
\end{array}
\right)
\end{align}
has characteristic polynomial
\begin{align}
&  \left(  1 + \frac{h^{(-1)}(\phi) h^{\prime\prime}\left(  h^{(-1)}%
(\phi)\right)  }{h'\left(  h^{(-1)}(\phi)\right)  } - \lambda
+\mathcal{O}(\phi^{n})\right)  (-\lambda+\mathcal{O}(\phi^{n})) - \underbrace{\frac{(\alpha+C
(\theta-2)) \theta}{2 C h^{(-1)}(\phi)^{2}}}_{\mathcal{O}(\phi^{(2n)}%
)}=0.\end{align}
That is, 
\begin{align}
&  \lambda\left(  \lambda-1-\frac{h^{(-1)}(\phi) h^{\prime\prime}\left(
h^{(-1)}(\phi)\right)  }{h'\left(  h^{(-1)}(\phi)\right)  }\right)
+\mathcal{O}(\phi^{n})=0.
\end{align}
with eigenvalues $\left\{  1+ \frac{h^{(-1)}(0) h^{\prime\prime}\left(
h^{(-1)}(0)\right)  }{h'\left(  h^{(-1)}(0)\right)  }, 0\right\}  $ as
$\phi\rightarrow0$.  Therefore, the line $L$ of equilibrium points is
normally hyperbolic. 

A set of non-isolated equilibrium points is said to be normally hyperbolic if the eigenvalues with zero real part correspond to eigenvectors which are tangent to the set. By definition, any point on a set of non-isolated singular points will have at least one eigenvalue zero. Then, all points in the set are non-hyperbolic. However, the stability of a normally hyperbolic set can be completely classified by considering the signs of eigenvalues in the remaining directions (i.e., for a curve, in the remaining $n-1$ directions) (see \cite{aulbach}, pp. 36). Therefore, the stability condition will be verified as  $\phi\rightarrow0$
is $\frac{h^{(-1)}(0) h^{\prime\prime}\left(  h^{(-1)}(0)\right)  }{h^{\prime
}\left(  h^{(-1)}(0)\right)  }<-1$. 

Setting, for example $h(u)=u^{-1/n}$, with
$n>1$, which satisfies the previous conditions 1, 2, and 3, we obtain
\begin{align}
&  \phi'=-\frac{\phi}{n} +\left(  \frac{1}{n}-\frac{\theta}{n}\right)
\phi^{n+1},\label{eq30}\\
&  \theta^{\prime  }= \phi^n\left(  \left(  1-\frac{\alpha}{2 C}\right)  \theta
-\frac{\theta^{2}}{2}\right)  . \label{eq31}%
\end{align}
The curve of equilibrium points $L: (\phi, \theta)= (0, \theta^{*})$ as
$\tau\rightarrow\infty$ for bounded $\theta$ has eigenvalues $\left\{
-\frac{1}{n},0\right\}  $, Therefore, it is normally hyperbolic and stable.

\subsection{Global dynamics}

Defining the compact variables
\begin{equation}
\Phi=\frac{2 \arctan(\phi)}{\pi}, \quad\Theta=\frac{2 \arctan(\theta)}%
{\pi},
\end{equation}
we obtain  
\begin{align}
&  \Phi'= \frac{\sin(\pi\Phi) \left(  -\left(  \tan\left(  \frac
{\pi\Theta}{2}\right)  -1\right)  \tan^{n}\left(  \frac{\pi\Phi}{2}\right)
-1\right)  }{\pi n},\label{eqPhi}\\
&  \Theta'= \frac{\tan^{n}\left(  \frac{\pi\Phi}{2}\right)  ((2
C-\alpha) \sin(\pi\Theta)+C (\cos(\pi\Theta)-1))}{2 \pi C}. \label{eqTheta}%
\end{align}

In this coordinates, the points with $\Phi=0$ correspond to $u\rightarrow
\infty$. The points with $\Phi=\pm1$ are representations of $u \rightarrow
0^{+}$ or $u \rightarrow0^{-}$, respectively. Moreover, $\Theta=\pm1$ are
representations of $\theta=1-u+v \rightarrow\pm\infty.$ As before, the
physical region corresponds to $\Phi\geq0$ (corresponding to $u\geq0$).

Finally, if $\theta$ is bounded as $\tau\rightarrow\infty$ (which we will have
so), we would have from (\ref{eq30}) and (\ref{eq31}) that
\begin{align}
&  \phi'=-\frac{\phi}{n} +\mathcal{O}(\phi^{n+1}),\label{asympt1}\\
&  \theta^{\prime  }= \phi^n \left(  \left(  1-\frac{\alpha}{2 C}\right)  \theta
-\frac{\theta^{2}}{2}\right)  +\mathcal{O}(\phi^{n+1}). \label{asympt2}%
\end{align}
\begin{figure}[ptb]
\centering
\includegraphics[width=0.8\textwidth]{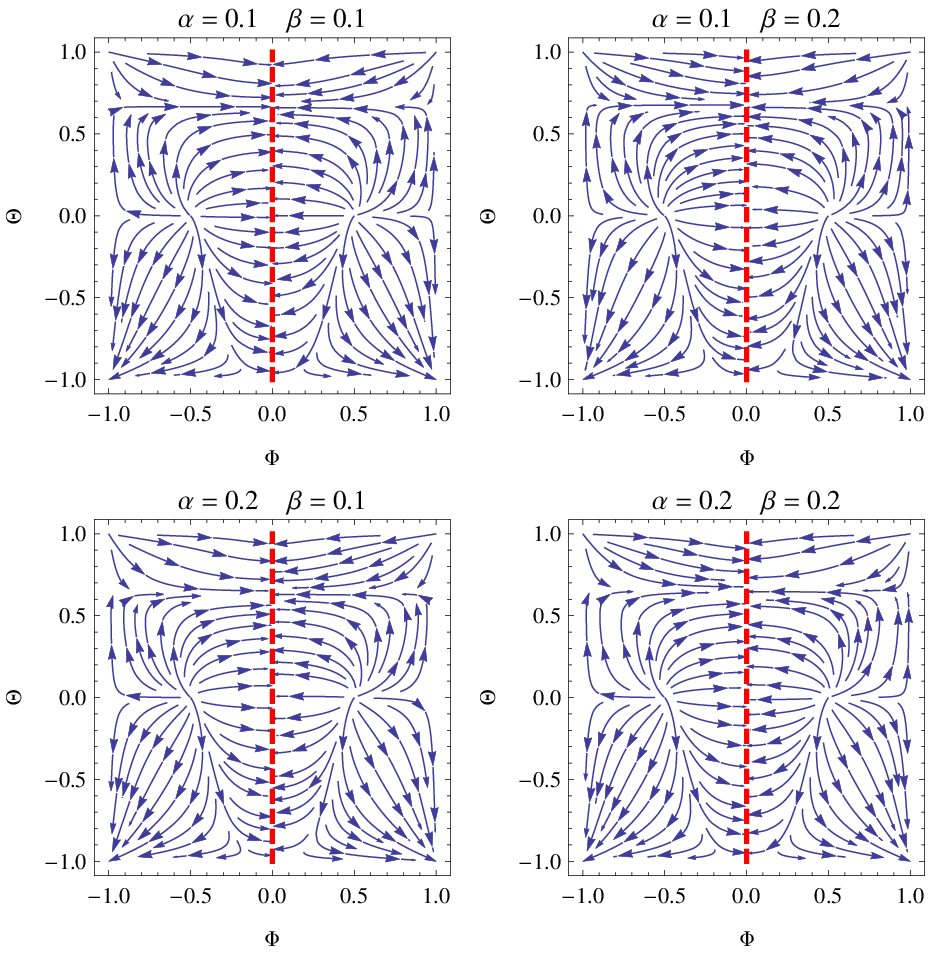} \caption{Phase plot of
(\ref{eqPhi}), (\ref{eqTheta}) for some $\alpha>0, \beta>0$, $n=2$, and the
choice $2 C:=\alpha+ \sqrt{\alpha^{2} + 4 \beta}$. The red dashed line is a
stable curve of equilibrium points $L: (\phi, \theta)= (0, \theta^{*})$. The physical
region is $\Phi\geq0$.}%
\label{fig:my_label4}%
\end{figure}\begin{figure}[ptb]
\centering
\includegraphics[width=0.8\textwidth]{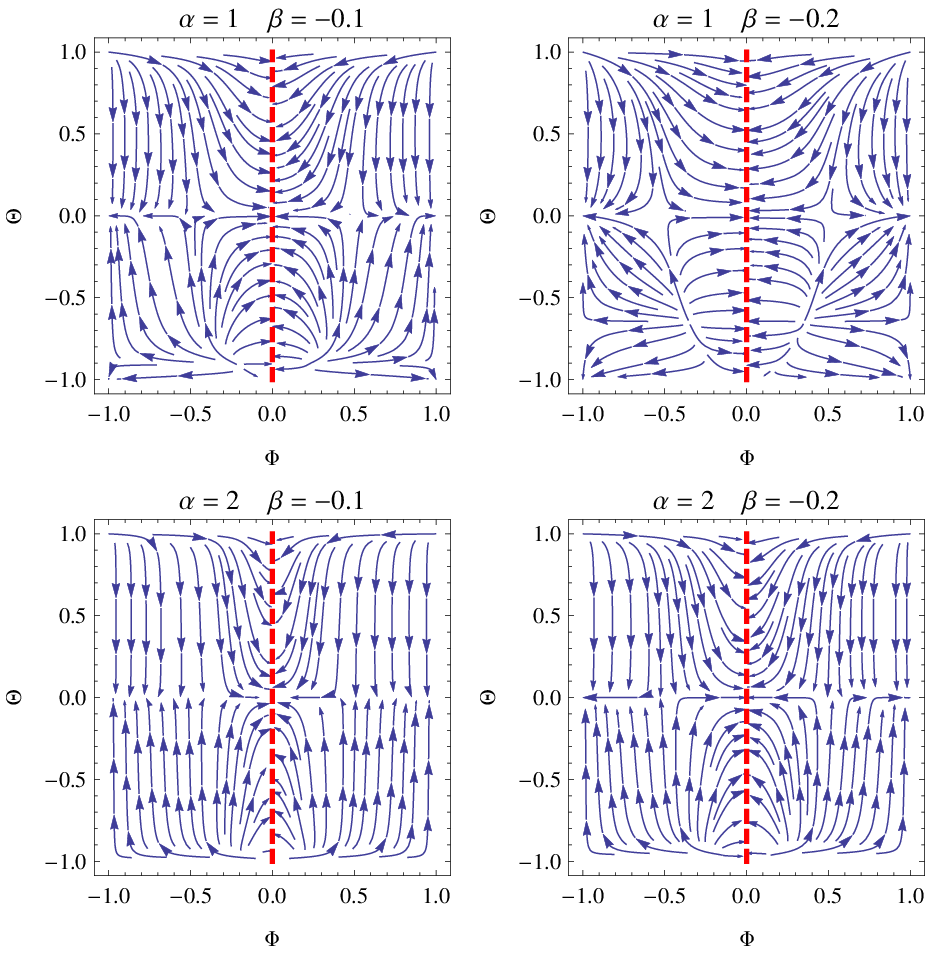} \caption{Phase plot of
(\ref{eqPhi}), (\ref{eqTheta}) for some $\alpha>0, \beta<0$ and the choice $2
C_{-}:=\alpha- \sqrt{\alpha^{2} - 4 |\beta|}$. The red dashed line is a stable
curve of equilibrium points $L: (\phi, \theta)= (0, \theta^{*})$. The physical region
is $\Phi\geq0$.}%
\label{fig:my_label5}%
\end{figure}\begin{figure}[ptb]
\centering
\includegraphics[width=0.8\textwidth]{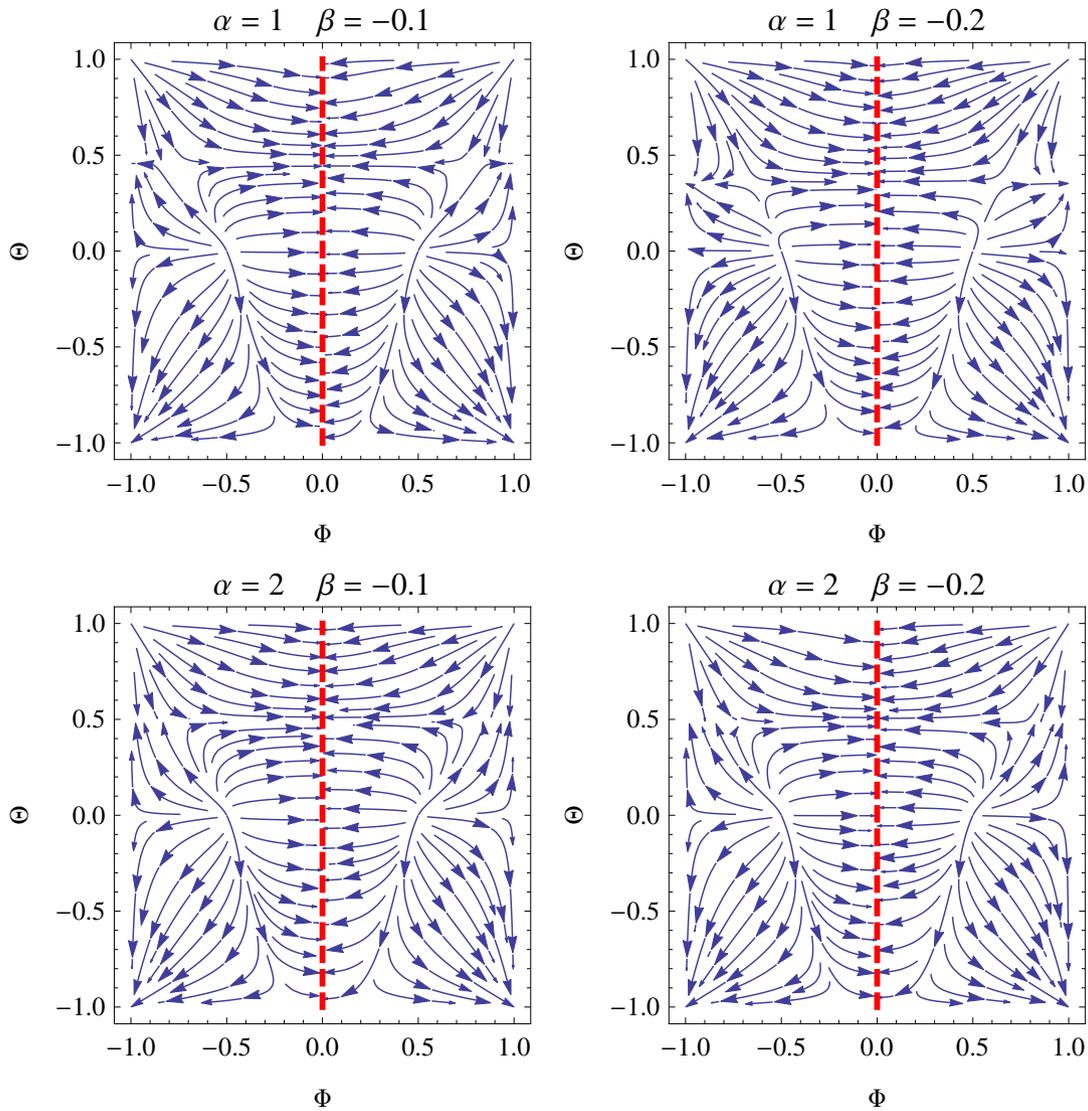} \caption{Phase plot of
(\ref{eqPhi}), (\ref{eqTheta}) for some $\alpha>0, \beta<0$ and the choice $2
C_{+}:=\alpha+ \sqrt{\alpha^{2} - 4 |\beta|}$. The red dashed line is a stable
curve of equilibrium points. The physical region is $\Phi\geq0$.}%
\label{fig:my_label6}%
\end{figure}The asymptotic equations (\ref{asympt1}), (\ref{asympt2}) as
$\phi\rightarrow0$ are integrable with solution
\begin{equation}
\left(
\begin{array}
[c]{c}%
\phi(\tau)\\
\theta(\tau)
\end{array}
\right)  = \left(
\begin{array}
[c]{c}%
e^{-\frac{\tau}{n}} c_{1}\\

\frac{2 C-\alpha}{C-\exp\left(  \frac{(2 C-\alpha) \left(  e^{-\tau} c_{1}%
^{n}+2 C c_{2}\right)  }{2 C}\right)  }%
\end{array}
\right)  ,
\end{equation}
converging to $L: (\phi, \theta)= (0, \theta^{*})$ as $\tau\rightarrow\infty$
for bounded $\theta$.

Figure \ref{fig:my_label4} shows the phase plot of system (\ref{eqPhi}),
(\ref{eqTheta}) for some $\alpha>0, \beta>0$, $n=2$, and the choice $2
C:=\alpha+ \sqrt{\alpha^{2} + 4 \beta}$. The red dashed line is a stable line
of equilibrium points $L: (\phi, \theta)= (0, \theta^{*})$. 

Figure \ref{fig:my_label5} shows the phase plot of system (\ref{eqPhi}),
(\ref{eqTheta}) for some $\alpha>0, \beta>0$, $n=2$, and the choice $2
C_{-}:=\alpha- \sqrt{\alpha^{2} - 4 |\beta|}$. The red dashed line is a stable
curve of equilibrium points $L: (\phi, \theta)= (0, \theta^{*})$. 

Figure \ref{fig:my_label6} shows the phase plot of system (\ref{eqPhi}),
(\ref{eqTheta}) for some $\alpha>0, \beta>0$, $n=2$, $\alpha^{2}+4 \beta\geq0$
and the choice $2 C_{+}:=\alpha+ \sqrt{\alpha^{2} - 4 |\beta|}$. The red
dashed line is a stable curve of equilibrium points $L: (\phi, \theta)= (0,
\theta^{*})$.

All these plots illustrate our analytical findings. These are: (i) The
solution of (\ref{bc}) $R = -Ct$ defined for $-\infty< t \leq0$, where $C>0$
is a fixed constant, is unstable. (ii) The curve of equilibrium points $L: (\phi,
\theta)= (0, \theta^{*})$ (i.e., $u\rightarrow\infty$, and $v\rightarrow
\infty$, in such a way that $1-u+v \rightarrow\theta^{*}$) is stable as
$\tau\rightarrow\infty$ for bounded $\theta$.

Result (ii) means that, as $\tau\rightarrow\infty$, we have 
\begin{equation}\label{eq47}
1- u(\tau ) +u'(\tau)= \theta^*, \quad u(\infty)=\infty. 
\end{equation}
The solution of (\ref{eq47}) is 
\begin{equation}
u(\tau)= c_1 e^{\tau }-\theta^*+1, \quad c_1 \neq 0.   
\end{equation}
Then, 
\begin{eqnarray}
&& \bar{R}(\tau)= R_s(\tau) u(\tau)= C e^{-\tau} u(\tau)  = C \left(c_1-(\theta^*-1) e^{-\tau }\right). 
\end{eqnarray}
Hence, 
\begin{equation}
R(t)=C \left(c_1+(\theta^*-1) t\right). \label{eq50}
\end{equation}
Substituting back in (\ref{bc}), we have that in order of (\ref{eq50}) to be an exact solution for (\ref{bc}) we must impose the compatibility condition: 
\begin{equation}
C \theta^* (\alpha +C (\theta^*-2))=0. 
\end{equation}
We have some specific solutions when 
$\theta^* \in \left\{0, 2-\frac{\alpha
}{C}\right\}$. 

However, recall that $\theta^*$ is an arbitrary constant value by definition of line $L$. So, the natural condition is
\begin{equation}
C=  \frac{\alpha }{2-\theta^*}. 
\end{equation}
Then,
the solution of (\ref{bc}) given by 
\begin{equation}\label{exact_R_1}
R(t)=   \frac{\alpha  c_1}{2-\theta^*}+\frac{\alpha  (\theta^*-1)    t}{2-\theta^*}, \quad c_1\neq 0,
\end{equation}
defined in the semi-infinite-interval
$-\infty< t\leq 0$, 
is stable as $t\rightarrow 0^{-}$ ($\tau \rightarrow +\infty$). Finally,
\begin{equation}
\lim_{t\rightarrow 0^{-}}  R(t) = \frac{\alpha  c_1}{2-\theta^*} \neq 0
\end{equation}
by construction. 

\section{Lie symmetries and singularity analysis}

\label{sec3}

In the following, we discuss the application of Lie's theory and the
singularity analysis for the derivation of exact and analytic solutions for
the master equation (\ref{bc}).

\subsection{Lie symmetries}

Consider the function $\Phi$ which describes the map of an one-parameter point
transformation such as $\Phi\left(  R\left(  t\right)  \right)  =R\left(
t\right)  \ $with infinitesimal transformation%
\begin{align}
t'  &  =t^{i}+\varepsilon\xi\left(  t,R\right), \label{sv.12}\\
R'  &  =R+\varepsilon\eta\left(  t,R\right)
\end{align}
and generator $\mathbf{X}=\frac{\partial t'}{\partial\varepsilon
}\partial_{t}+\frac{\partial x'}{\partial\varepsilon}\partial_{R}%
,~$where~{$\varepsilon$ is the parameter of smallness}; $t$ is the independent
variable and $R\left(  t\right)  $ the dependent variable.

Assume $R\left(  t\right)  $ be a solution for the differential equation
$\mathcal{H}\left(  t,R,\dot{R},\ddot{R}\right)  =0.$ Then, under the
one-parameter map $\Phi$, function $R'\left(  t'\right)
=\Phi\left(  R\left(  t\right)  \right)  $ is a solution for the differential
equation $\mathcal{H}$, if and only if the differential equation is also
invariant under the action of the map, $\Phi$, that is, the following
condition holds  \cite{kumei}
\begin{equation}
\Phi\left(  \mathcal{H}\left(  t,R,\dot{R},\ddot{R}\right)  \right)
=0.\label{f1}%
\end{equation}

For every map $\Phi$ in which condition (\ref{f1}) holds, means that the
generator~$X$ is a Lie point symmetry for the differential equation, while the
following condition is true
\begin{equation}
\mathbf{X}^{\left[  2\right]  }\left(  \mathcal{H}\left(  t,R,\dot{R},\ddot
{R}\right)  \right)  =0. \label{sv.17}%
\end{equation}
Vector field $\mathbf{X}^{\left[  1\right]  }$ describes the first extension
of the symmetry vector in the jet-space of variables, $\left\{  t,R,\dot
{R},\ddot{R}\right\}  $ defined as%
\begin{equation}
X^{\left[  2\right]  }=X+\eta^{\left[  1\right]  }\partial_{\dot{R}}%
+\eta^{\left[  2\right]  }\partial_{\ddot{R}},
\end{equation}
with $\eta^{\left[  1\right]  }=\dot{\eta}-\dot{R}\xi$ and $\eta^{\left[
2\right]  }=\dot{\eta}^{\left[  1\right]  }-\ddot{R}\dot{\xi}$.

The existence of a Lie symmetry for a given differential equation indicates
the associated Lagrange's system, $\frac{dt}{\xi}=\frac{dR}{\eta},~$in which
the solution of this system provides the invariant functions which can be used
to reduce the order of the ordinary differential equation.

Let us now turn our attention to the boundary condition (\ref{bc}). By
applying Lie's theory, it follows that Equation (\ref{bc}) is invariant under
the infinitesimal transformation  \cite{kumei}%
\begin{align}
\overline{t} &  \rightarrow t+\varepsilon\left(  \alpha_{1}+\alpha_{2}t\right),\\
\bar{R} &  \rightarrow R+\varepsilon\left(  \alpha_{2}R\right),
\end{align}
where $\varepsilon$ is the infinitesimal parameter, that is, $\varepsilon
^{2}\simeq0.~$

Hence, equation (\ref{bc}) admits as Lie point symmetries the vector fields
$X_{1}=\partial_{t}$ and $X_{2}=t\partial_{t}+R\partial_{R}$, with commutator
$\left[  X_{1},X_{2}\right]  =X_{1}$. The Lie symmetries $\left\{  X_{1}%
,X_{2}\right\}  $ form the $A_{2,2}$ Lie algebra in the Morozov-Mubarakzyanov
classification scheme  \cite{mb1}. We continue by applying the corresponding
Lie invariants to reduce the order of the master equation (\ref{bc}).

From the Lie point symmetry $X_{1}$ we define the differential
invariants$~y=\dot{R}~,~x=R$. Thus, by assuming $x$ to be the new independent
variable and $y=y\left(  x\right)  $, equation (\ref{bc}) is written as
\begin{equation}
2xy\frac{dy\left(  x\right)  }{dx}+y^{2}\left(  x\right)  +\alpha y\left(
x\right)  -\beta=0~. \label{bcc1}%
\end{equation}
Equation (\ref{bcc1}) admits the Lie symmetry vector $\Gamma^{1}=\left(
y\left(  x\right)  +\alpha y-\frac{\beta}{y\left(  x\right)  }\right)
\partial_{y}$ which provides the Lie's integration factor $\mu=\left(
2x\left(  y^{2}\left(  x\right)  +\alpha y-\beta\right)  \right)  ^{-1}$;
hence, by multiply equation (\ref{bcc1}) with it we end with the equation
\begin{equation}
\int\frac{y}{\left(  y^{2} +\alpha y-\beta\right)  }%
dy=-\frac{dx}{2x},
\end{equation}
that is,%
\begin{equation}
\ln\left(  x-x_{0}\right)  =-\ln\left(  y^{2}+\alpha y-\beta\right)
-\frac{2\alpha}{\sqrt{\alpha^{2}+4\beta}}\text{arctanh}\left(  \frac{2y+\alpha
}{\sqrt{\alpha^{2}+4\beta}}\right)  . \label{bcc2}%
\end{equation}

When $\beta=-\frac{\alpha^{2}}{4}$, solution (\ref{bcc2}) is written
\begin{equation}
\ln\left(  2+y+\alpha\right)  +\frac{\alpha}{2y+\alpha}=-\frac{1}{2}\ln\left(
x-x_{0}\right)  .
\end{equation}

Moreover, application of the symmetry vector $X_{2}$ in (\ref{bc}) provides
the first-order ordinary differential equation%
\begin{equation}
2z\left(  y\left(  z\right)  -z\right)  \frac{dy\left(  z\right)  }{dz}%
+y^{2}\left(  z\right)  +\alpha y\left(  z\right)  -\beta=0. \label{bcc3}%
\end{equation}
The latter equation belongs to the family of Abel equations of the second
kind. Equation (\ref{bcc3}) can be integrated similarly with equation
(\ref{bcc1}) with the derivation of Lie's integration factor. For a constant
value $y=y_{0}$, with $y_{0}^{2}+\alpha y_{0}-\beta=0$ it is clear that the
exact solution of  \cite{bhui} is recovered.

\subsection{Singularity analysis\newline}

The ARS algorithm  \cite{ars1,ars2,ars3} has three basic steps which we briefly
discuss. The first step is based on the determination of \ the
leading-order behavior, at least in terms of the dominated exponent. The
coefficient of the leading-order term may or may not be explicit. This
indicates the existence of a movable singularity for the given differential
equation. A second step is the determination of the exponents at which the
arbitrary constants of integration enter. This step provides information about
the existence of integration constants for the differential equation. Finally,
the third step is called the consistency test, where we substitute an expansion
up to the maximum resonance into the full equation to check if solves the equation.

For the singularity analysis to work the exponents of the
the leading-order term needs to be a negative integer or a nonintegral rational
number, while the resonances should be rational numbers, while one of the
resonances should be $-1\,\ $which warranty the singularity is a movable pole.
Excluding the generic resonance $-1$, the analytic solution is expressed by a
Right Painlev\'{e} Series if the rest of the resonances are nonnegative, for a
Left Painlev\'{e} Series the resonances must be nonpositive while for a full
Laurent expansion the resonances they have to be mixed. Clearly for a
second-order ordinary equation \ the possible Laurent expansions are Left or
Right Painlev\'{e} Series.

We apply the ARS algorithm for equation (\ref{bc}), from the fist stem we
determine\ the leading-order term to be $R_{\text{leading}}\left(  t\right)
=R_{0}\left(  t-t_{0}\right)  ^{\frac{2}{3}}$, where $t_{0}$ indicates the
location of the movable singularity and $R_{0}$ is arbitrary. From the second
step, the resonances, are derived to be $s_{1}=-1$ and $s_{4}=4$, which means
that the analytic solution of (\ref{bc}) can be expressed in terms of the
Right Painlev\'{e} Series%
\begin{equation}
R\left(  t\right)  =R_{0}\left(  t-t_{0}\right)  ^{\frac{2}{3}}+R_{1}\left(
t-t_{0}\right)  +R_{2}\left(  t-t_{0}\right)  ^{\frac{4}{3}}+R_{3}\left(
t-t_{0}\right)  ^{\frac{5}{3}}+\ldots~. \label{Solution_2}
\end{equation}
We replace in (\ref{bc}) from where we find that
\begin{equation}
R_{1}=-\frac{3\alpha}{4},~R_{2}=\frac{9}{320R_{0}}\left(  3\alpha^{2}%
+16\beta\right),~R_{2}=\frac{3\alpha}{320R_{0}^{2}}\left(  3\alpha
^{2}+16\beta\right),\ldots.%
\end{equation}

In the special case in which $\left(  3\alpha^{2}+16\beta\right)  =0$, that is
$\beta=-\frac{3\alpha^{2}}{16}$, we find that $R_{I}=0,~I>1$, thus we end with
the closed-form solution%
\begin{equation}\label{closed-form-solution}
R_{s}\left(  t\right)  =R_{0}\left(  t-t_{0}\right)  ^{\frac{2}{3}}%
-\frac{3\alpha}{4}\left(  t-t_{0}\right), \quad \beta=-\frac{3\alpha^{2}}{16},%
\end{equation}
which can be seen as extension of the exact solution of  \cite{bhui}. Indeed
for large values of $t-t_{0}$, $R_{s}\left(  t\right)  \simeq-\frac{3\alpha
}{4}\left(  t-t_{0}\right)  $, however, for small values of $\left(
t-t_{0}\right)  $, the term $\left(  t-t_{0}\right)  ^{\frac{2}{3}}$ dominates
such that $R_{s}\left(  t\right)  \simeq R_{0}\left(  t-t_{0}\right)
^{\frac{2}{3}}$.

It is clear that from the symmetry analysis and the singularity analysis
new exact solutions were found for the master equation (\ref{bc}).

\section{Conclusions}
\label{sec4}

We performed a detailed study for the master equation of the temporal equation
of radiating stars by investigating the global dynamics, the Lie point
symmetries and applying the ARS algorithm of singularity analysis. From the
analysis of the global dynamics of the master equation we were able to find  a
new asymptotic behavior which corresponds to a new exact solution for a
radiating star spacetime, as also to extract important information to
infer about the stability and the physical properties of the asymptotic solutions.

We have new analytical findings. The
solution of (\ref{bc}) $R = -Ct$ defined for $-\infty< t \leq0$, where $c_1>0$
is a fixed constant, is unstable. The curve of equilibrium points $L: (\phi,
\theta)= (0, \theta^{*})$ has associated a family of solutions of (\ref{bc}) given by \eqref{exact_R_1}, 
defined in the semi-infinite-interval
$-\infty< t\leq 0$, 
which is stable as $t\rightarrow 0^{-}$ ($\tau \rightarrow +\infty$). 
Furthermore, the Lie symmetry analysis provides that the master equation
admits two Lie point symmetries which form the $A_{2,2}$ Lie algebra. Each Lie
symmetry was applied to reduce the order for the ordinary
differential equation into a first-order differential equation, while new
exact similarity solutions were found.

Finally, from the application of the ARS algorithm, we found that the master
equation possesses the Painlev\'{e} property and we were able to write the
analytic solution for the radiating stars in order of Painlev\'{e} Series \eqref{Solution_2}, 
where $R_0$ is arbitrary, $R_1$ is function of $\alpha$, and the $R_j, j\geq 2$ are functions of parameters $\alpha$ and $\beta$. 
Additionally, we have obtained the closed-form solution \eqref{closed-form-solution} for $\beta=-\frac{3\alpha^{2}}{16}$. 
To summarise, we have investigated the global dynamics for the given master ordinary differential equation to understand the evolution of solutions for various initial conditions as also to investigate the existence of asymptotic solutions. Moreover, with the application of Lie's theory, we have reduced the order of the master differential equation, while an exact similarity solution was determined. Finally, the master equation possesses the Painlev\'{e} property. Therefore, the analytic solution can be expressed in terms of a Laurent expansion. Such analyses are essential for understanding the physical properties of spacetimes which describe radiating stars.

\section*{Acknowledgements}
AP \& GL were funded by Agencia Nacional de Investigaci\'{o}n y Desarrollo -
ANID through the program FONDECYT Iniciaci\'{o}n grant no. 11180126.
Additionally, GL thanks the support of Vicerrector\'{\i}a de Investigaci\'{o}n
y Desarrollo Tecnol\'{o}gico at Universidad Cat\'{o}lica del Norte.

%\section*{Author Contributions}

%The authors equally contributed. 

\section*{CONFLICT OF INTEREST}

The authors declare to have no conflict of interest.

\section*{ORCID}

\noindent Genly Leon\orcidA{} \url{https://orcid.org/0000-0002-1152-6548}

\noindent Megandhren Govender\orcidB{} \url{https://orcid.org/0000-0001-6110-9526}

\noindent Andronikos Paliathanasis\orcidC{} \url{https://orcid.org/0000-0002-9966-5517}

\end{document}